\documentclass[12pt]{iopart}
\usepackage{iopams}
\newcommand{\re}{\mathop{\rm Re}\nolimits}
\newcommand{\im}{\mathop{\rm Im}\nolimits}
\newcommand{\al}{\alpha}
\newcommand{\be}{\beta}
\newcommand{\ep}{\epsilon}
\newcommand{\de}{\delta}
\newcommand{\ze}{\zeta}

\newcommand{\CC}{{\mathbb C}}
\newcommand{\NN}{{\mathbb N}}
\newcommand{\ZZ}{{\mathbb Z}}
\newcommand{\RR}{{\mathbb R}}
\newcommand{\cB}{{\mathcal B}}
\newcommand{\cH}{{\mathcal H}}
\newcommand{\cM}{{\mathcal M}}
\newcommand{\cN}{{\mathcal N}}
\newcommand{\cP}{{\mathcal P}}
\newcommand{\cS}{{\mathcal S}}
\newcommand{\fsl}{\mathfrak{sl}}
\newcommand{\pa}{\partial}
\newcommand{\id}{1\hspace{-.25em}{\rm I}}
\newcommand{\nid}{\noindent}
\newtheorem{theorem}{\bf Theorem}

\begin{document}
\jl{1}
\title[QES models in nonlinear optics]{Quasi-exactly solvable models
       in nonlinear optics}
\author{G \'Alvarez,
        F Finkel,
        A Gonz\'alez-L\'opez and
        M A Rodr\'{\i}guez}
\address{Departamento de F\'{\i}sica Te\'orica II,
         Facultad de Ciencias F\'{\i}sicas,
         Universidad Complutense, 28040 Madrid, Spain}
\begin{abstract}
We study a large class of models with an arbitrary (finite) number of
degrees of freedom, described by Hamiltonians which are polynomial in
bosonic creation and annihilation operators, and including as particular
cases $n$-th harmonic generation and photon cascades.  For each model,
we construct a complete set of commuting integrals of motion of the
Hamiltonian, fully characterize the common eigenspaces of the
integrals of motion, and show that the action of the Hamiltonian in
these common eigenspaces can be represented by a quasi-exactly solvable
reduced Hamiltonian, whose expression in terms of the usual generators
of $\fsl_{2}$ is computed explicitly.
\end{abstract}
\submitted
\maketitle
\section{Introduction}
Dating back at least to 1990, when Zaslavskii~\cite{Za90} pointed out
the relation between some quasi-exactly solvable (QES)
systems~\cite{ZU84,ZU87,Tu88,Sh89,Us94,GKO94} and the
Dicke and Heisenberg models, there has been an ongoing interest in
the correspondence between the exactly calculable part of the spectrum
of QES systems and some finite-dimensional systems like the spin systems
discussed by Zaslavskii or the effective Hamiltonians frequently used
in nonlinear optics.

In essence, Zaslavskii's method consisted in making a suitable ansatz
for the coefficients of the eigenvectors of each ``multiplet'' of the
spin system, coefficients which could be grouped into a generating function
that turned out to be related by elementary transformations to the
wavefunction of the QES system.

In 1995, \'Alvarez and \'Alvarez-Estrada~\cite{AA95} studied the usual
second harmonic generation effective Hamiltonian by showing its
equivalence to an infinite family of one-dimensional QES systems.  In
their method, the transition from the ``discrete'' photon system to
the ``continuous'' QES systems is achieved by first transforming the
effective Hamiltonian from the second quantization to the Bargmann
representation.  The key idea to separate variables and identify the
one-dimensional QES systems---the second step of their
procedure---is to use as a variable an appropriate quotient of powers
of the Bargmann variables that describe each oscillator.  (Later, the
same authors~\cite{AA01} used this method to study the problem of
third harmonic generation and pointed out that it works in general
for the $n$-th harmonic generation.)

Two recent papers by Dolya and Zaslavskii~\cite{DZ00,DZ01} proceed in
a related but different direction: they study models with a single
degree of freedom whose Hamiltonians are even polynomials in the
creation and annihilation operators, and show that under certain
conditions these Hamiltonians also lead to a QES system.

Finally, an important idea for our work can be traced back to the
early papers on QES models by Ushveridze and Zaslavskii, and appears
with more or less emphasis in~\cite{Za90,AA95,Us94,DZ01,AA01}: the
role played by integrals of motion to reduce completely integrable
Hamiltonians with more than one degree of freedom to a family of QES
systems.

In this paper we study a large class of models with an arbitrary
(finite) number of degrees of freedom described by Hamiltonians which
are polynomial in the creation and annihilation operators.  These
models include as particular cases the effective Hamiltonians of
$n$-th harmonic generation~\cite{BM00,KS00}, as well as photon
cascades of equal or different frequencies~\cite{KGV01}, which, to the
best of the authors' knowledge, have not been analyzed before as QES
systems.  For each model, we first construct a complete set of
commuting integrals of motion of the Hamiltonian.  We use the Bargmann
representation to completely characterize the common eigenspaces of
the integrals of motion, which are of course invariant under the
Hamiltonian.  We show that the action of the Hamiltonian in these
common eigenspaces can be represented by a QES reduced Hamiltonian,
whose expression in terms of the usual generators of $\fsl_{2}$ is
computed explicitly.  We emphasize that the derivation of this
explicit expression is not just an exercise of academic interest but
the starting point for the application of asymptotic methods to
calculate the corresponding eigenvalues in the limit of a large number
of photons.  By way of example, we give explicitly the expressions for
$n$-th harmonic generation and $N$-photon cascades, which in the
former case can be compared with the particular instances $n=2,3$
studied in~\cite{AA95,AA01}. The paper ends with a brief summary.

\section{The models}
In this paper we shall study the following general Hamiltonian
\begin{equation}
  H = H_0 + g H_1
  \label{sys}
\end{equation}
where
\begin{eqnarray}
  H_0 = \sum_{l=1}^{N} \nu_l a_l^{\dagger}a_l
        +
        \sum_{k=1}^{M} \mu_k b_k^{\dagger}b_k
        \qquad \nu_l,\mu_k>0 \qquad N,M\in\NN \label{H0}\\
  H_1 = \prod_{k=1}^M (b_k^{\dagger})^{m_k}
        \cdot
        \prod_{l=1}^N a_l^{n_l}
        +
        \prod_{l=1}^N (a_l^{\dagger})^{n_l}
        \cdot
        \prod_{k=1}^M b_k^{m_k} \qquad n_l,m_k\in \NN \label{H1}
\end{eqnarray}
and the frequencies $\nu_l$ and $\mu_k$ are subject to the
energy-conservation constraint
\begin{equation}
  \sum_{l=1}^N n_l\nu_l=\sum_{k=1}^M m_k\mu_k\,.
  \label{constraint}
\end{equation}
In the preceding expressions, the operators $a_l,b_k$ ($a^{\dagger}_l,
b^{\dagger}_k$) represent boson annihilation (creation) operators
with frequencies $\nu_l,\mu_k$, respectively.
The Hamiltonian~(\ref{sys}) is an effective model describing the
conversion of a number of photons of frequencies
$\nu_1,\ldots,\nu_N$ into photons of frequencies
$\mu_1,\ldots,\mu_M$. It includes as particular cases the usual
processes of $n$-th harmonic generation (if $N=M=m_1=1$, $n_1=n$)
and multiple photon cascades (if $M=m_1=n_1=\cdots = n_N=1$),
which have recently received considerable attention in the
literature~\cite{BM00, KS00,KGV01}.

A key property common to all these models is that they are
\emph{completely integrable} in the sense that there exists a set
of $N+M$ (the number of degrees of freedom) pairwise commuting,
functionally independent operators including the Hamiltonian.
These sets are clearly not unique.  We have found it convenient to
take as our set
\begin{equation}
  \{H_0,H_1,A_l,B_k\mid l=1,\dots,N-1;\;k=1,\ldots,M-1\}
  \label{commset}
\end{equation}
which includes the unperturbed Hamiltonian $H_{0}$ and the perturbation
$H_{1}$ separately, and where
\begin{eqnarray}
  A_l = n_{l+1}a_{l}^{\dagger}a_{l}-n_{l}a_{l+1}^{\dagger}a_{l+1}
        \label{opa} \\
  B_k = m_{k+1}b_{k}^{\dagger}b_{k}-m_{k}b_{k+1}^{\dagger}b_{k+1}.
        \label{opb}
\end{eqnarray}
A straightforward computation using the standard bosonic commutation
relations shows that these operators commute pairwise, and we shall
see below that they are in fact functionally independent.  As a
consequence, there exists a basis of common eigenfunctions of all the
operators in the set.  We will see in the next section that it is
possible to give an explicit description of the simultaneous
eigenspaces of $H_{0}$, $A_{l}$ and $B_{k}$ (the ``unperturbed
eigenspaces'').  By studying the action of $H_{1}$ within these
subspaces, in the next section we shall establish the exact solvability
of the model~(\ref{sys})--(\ref{constraint}) and derive some general
properties of its spectrum.
\section{Invariant subspaces and matrix representation}

Following~\cite{AA95}, we shall study the spectrum of the
Hamiltonian~(\ref{sys}) using the Bargmann
representation~\cite{GP90}. In this representation, the Hilbert
space for a single boson is the space of entire functions of the
form
\begin{equation}
  f(z) = \sum_{n=0}^{\infty} \frac{c_{n}}{\sqrt{n!}}\,z^n
  \qquad z\in\CC
\end{equation}
where the complex numbers $c_{n}$ are such that
\begin{equation}
  \sum_{n=0}^{\infty} |c_{n}|^2 < \infty
\end{equation}
and the scalar product is defined by
\begin{equation}
  (g,\,f)
  =
  \frac{1}{\pi}
  \int_{\RR\times\RR}\rmd(\re z)\,\rmd(\im z)\,
    \overline{g(z)} f(z) \rme^{-|z|^2}.
  \label{eq:sp}
\end{equation}
The orthonormal harmonic oscillator eigenstates
$|n\rangle$ are given in this representation by
\begin{equation}
  |n\rangle \rightarrow \frac{z^n}{\sqrt{n!}}
  \qquad
  n = 0, 1, \ldots
  \label{eq:hoe}
\end{equation}
and the annihilation and creation operators are represented by
derivation with respect to $z$ and multiplication by $z$, respectively
\begin{equation}
  a \rightarrow \frac{\rmd}{\rmd z}
  \qquad
  a^{\dagger} \rightarrow z.
\end{equation}
The Hilbert space $\cH$ of the model~(\ref{sys}) is the tensor
product of $N+M$ single boson Hilbert spaces. Denoting by $x_{l}$
($l=1,\ldots, N$) and $y_{k}$ ($k=1,\ldots, M$) the complex
variables associated to the ``$a$'' and ``$b$'' degrees of
freedom, respectively, we have the following assignments:
\begin{equation}
  a_l\rightarrow \partial_{x_l}
  \qquad
  a_l^{\dagger}\rightarrow x_l
  \qquad
  b_k\rightarrow \partial_{y_k}
  \qquad
  b_k^{\dagger}\rightarrow y_k.
\end{equation}
A basis of the Hilbert space $\cH$ of our models is the set of
(in general unnormalized) monomials
\begin{equation}
  x^iy^j \qquad i\in\ZZ_+^N,\,\,j\in\ZZ_+^M
  \label{monomial}
\end{equation}
where $\ZZ_+$ denotes the set of nonnegative integers,
and we have used the multiindex notation
\begin{equation}\label{multi1}
  x^i \equiv \prod_{l=1}^N x_l^{i_l} \qquad
  y^j \equiv \prod_{k=1}^M y_k^{j_k}.
\end{equation}
The corresponding expressions for the operators in the
set~(\ref{commset}) are
\begin{eqnarray}
  H_0 =   \sum_{l=1}^{N}\nu_l x_l\pa_{x_l}
        + \sum_{k=1}^{M}\mu_k y_k\pa_{y_k} \label{oph0}\\
  H_1 = y^m\pa_x^n + x^n\pa_y^m \label{oph1}\\
  A_l = n_{l+1}x_{l}\pa_{x_l} - n_{l}x_{l+1}\pa_{x_{l+1}} \label{opal}\\
  B_k = m_{k+1}y_{k}\pa_{y_k} - m_{k}y_{k+1}\pa_{y_{k+1}} \label{opbk}
\end{eqnarray}
where $x^n$ and $y^m$ are monomials as in equation~(\ref{multi1}) and
\begin{equation}
  \pa_x^n \equiv \prod_{l=1}^N\pa_{x_l}^{n_l}
  \qquad
  \pa_y^m \equiv \prod_{k=1}^M\pa_{y_k}^{m_k}.
  \label{multi2}
\end{equation}
(In the rest of the paper we shall make frequent use of the multiindex
notation without further notice whenever there are no ambiguities in
the interpretation of the formulae.)

We first note that each monomial~(\ref{monomial}) is a common eigenfunction
of the operators $H_0$, $A_l$, $B_k$, $l=1,\ldots,N-1$, $k=1,\ldots,M-1$,
with eigenvalues respectively given by
\begin{eqnarray}
  E_0 = \sum_{l=1}^N \nu_l i_l+\sum_{k=1}^M \mu_k j_k \label{E0}\\
  \al_l = n_{l+1}i_l-n_l i_{l+1} \qquad l=1,\ldots,N-1\\
  \be_k = m_{k+1}j_k-m_{k}j_{k+1}\qquad k=1,\ldots,M-1.\label{beta}
\end{eqnarray}
The spectrum of the Hamiltonian~(\ref{sys}) can thus be computed
by diagonalizing the perturbation Hamiltonian $H_1$ in the
common eigenspaces
$\cS_{E_0,\al_1,\ldots,\al_{N-1},\be_1,\ldots,\be_{M-1}}$, spanned
by the monomials~(\ref{monomial}) whose exponents
satisfy equations~(\ref{E0})--(\ref{beta}) for fixed values of
$E_0,\al_1,\ldots,\be_{M-1}$.

We shall now describe the common eigenspaces
$\cS_{E_0,\al_1,\ldots,\al_{N-1},\be_1,\ldots,\be_{M-1}}\equiv\cS$.
Since the unperturbed energy $E_0$ is a finite sum of nonnegative terms,
it follows that $\cS$ is finite-dimensional.  Thus,
the model~(\ref{sys}) is \emph{exactly solvable}, since its whole
spectrum can be computed algebraically.

Let $x^p y^q$ be a given (fixed) monomial in $\cS$.  We shall first
prove that $\cS$ is spanned by monomials of the form
\begin{equation}
  f_s = x^p y^q \ze^s
  \label{fs}
\end{equation}
where
\begin{equation}
  \ze = \frac{x^n}{y^m}
  \label{zeta}
\end{equation}
and $s$ ranges over a finite interval $s_{0}\le s\le s_{1}$ of $\ZZ$.

Indeed, let $x^{p'}y^{q'}$ be any other monomial in $\cS$. The
quotient $Q=x^{p'-p}y^{q'-q}$ is then a \emph{generalized} monomial
(i.e. a monomial with possibly negative integer exponents) and thus
may not belong to $\cH$. However, the action of the
operators~(\ref{oph0})--(\ref{opbk}) on such generalized
monomials is well-defined, and a simple calculation shows that $Q$
satisfies the system of first-order partial differential equations
\begin{equation}
  H_0Q = A_1Q = \cdots = A_{N-1}Q = B_1Q = \cdots = B_{M-1}Q = 0\,.
  \label{system}
\end{equation}
In other words, $Q$ must be a joint invariant of the $M+N-1$ commuting
vector fields $H_0$, $A_l$, $B_k$.  A second
straightforward computation shows that the function $\ze$ defined
in~(\ref{zeta}) is a joint invariant of this set of vector fields.
Since the number of independent variables is $N+M$, the general
solution of the system~(\ref{system}) is an arbitrary (smooth)
function of $\ze$, and since $Q$ is a generalized monomial, it must be
a power of $\ze$, which proves the first part of our claim.

Since $\cS$ is finite-dimensional, the values of $s$ for which
$f_s$ lies in $\cS$ must be bounded above and below.
If
\begin{equation}
\label{bounds}
s_0=\min\{s\in \ZZ: f_s\in \cS\}\\
s_1=\max\{s\in \ZZ: f_s\in \cS\}
\end{equation}
we shall next prove that
\begin{equation}
\cS=\langle f_s\,\mid\, s_0\le s\le s_1\rangle
\end{equation}
where $\langle\:\rangle$ denotes the linear span.
Indeed, since clearly
\begin{equation}
H_0f_s=E_0f_s\\
A_lf_s=\al_lf_s\\
B_kf_s=\be_kf_s
\end{equation}
for all integer values of $s$, it suffices to verify that
for all $s\in[s_0,s_1]\cap\ZZ$ the exponents of $f_s$ are all nonnegative integers.
Since $s\in[s_0,s_1]$ and both $f_{s_0}$ and $f_{s_1}$ belong to $\cS$,
this follows from the inequalities
\begin{equation}\label{inequalities}
p_l+n_l s\ge p_l+ n_l s_0 \ge 0\\
q_k-m_k s\ge q_k- m_k s_1 \ge 0\,.
\end{equation}
This completes the proof of our claim.

Note that (\ref{bounds}) and the inequalities (\ref{inequalities}) imply
that the bounds $s_0$ and $s_1$ are given by
\begin{equation}\fl
s_0=\max\left\{-\left[{p_l\over n_l}\right]: 1\le l\le N\right\}\qquad
s_1=\min\left\{\left[{q_k\over m_k}\right]: 1\le k\le M\right\}
\end{equation}
where $[\cdot]$ denotes the integer part.
Calling
\begin{equation}
  \cN_l = p_l+s_0 n_l\ge 0
          \qquad
      \cM_k=q_k-s_0 m_k\ge 0
\end{equation}
the eigenspace $\cS$ can be alternatively written as
\begin{equation}
  \cS \equiv \cS^{\cN}_{\cM}
  =
  x^{\cN}y^{\cM} \langle 1,\zeta,\ldots, \zeta^r\rangle
  \label{mod}
\end{equation}
where
\begin{equation}
  r = \min\left\{\left[{\cM_k\over m_k}\right]: 1\le k\le M\right\}
  \label{r}
\end{equation}
and the nonnegative integers $\cN_l$ and $\cM_k$
are subject to the single restriction
\begin{equation}
   \cN_l<n_l \quad \mbox{for al least one}
             \quad l\in\{1,\dots,N\}
  \label{restric}
\end{equation}
which is an immediate consequence of the definition of $s_0$. The
set of $N+M$ nonnegative integers
$\{\cN_1,\dots,\cN_N,\cM_1,\dots,\cM_M\}$
subject to the condition~(\ref{restric}) define uniquely the
eigenspace $\cS$. Indeed, $\cN_l$ is the minimum power of $x_l$
and $\cM_k$ is the maximum power of $y_k$ of the monomials in $\cS$.
We have thus proved the main result of this section:
\begin{theorem}
The common eigenspaces of the operators $H_0$, $A_l$, $B_k$,
$l=1,\dots,N-1$, $k=1,\dots,M-1$, are the spaces
$\cS^{\cN}_{\cM}$ given
in equations~\emph{(\ref{mod})--(\ref{restric})}.
\end{theorem}
Since these eigenspaces $\cS^{\cN}_{\cM}$ are invariant under $H_{1}$
(because this operator commutes with $H_0$ and $A_l$, $B_k$ for
all $l,k$), we devote the next paragraphs to study the
corresponding action.

Consider the basis $\cB$ of $\cS^{\cN}_{\cM}$ spanned by the normalized
vectors
\begin{equation}\label{basis}
  e_s = {1\over \sqrt{C_s}}\,x^{\cN+s n}y^{\cM-s m}
\qquad s=0,1,\ldots,r
\end{equation}
where~(see~equation~(\ref{eq:hoe}))
\begin{equation}\label{norm}
  C_s = \prod_{l=1}^N(\cN_l+n_l s)!\cdot\prod_{k=1}^M(\cM_k-m_k s)!\,.
\end{equation}
A straightforward computation using equations~(\ref{oph1}), (\ref{basis})
and~(\ref{norm}) yields
\begin{eqnarray}
\fl
  H_1 e_s
  =
  \sqrt{C_{s-1}\over C_s}
  \prod_{l=1}^N(\cN_l+n_l s)\cdots(\cN_l+n_l(s-1)+1)\,e_{s-1}\nonumber\\
  {}+
  \sqrt{C_{s+1}\over C_s}
  \prod_{k=1}^M(\cM_k-m_k s)\cdots(\cM_k-m_k(s+1)+1)\,e_{s+1}\,.
\end{eqnarray}
Thus, the matrix $\sf H_1$ representing $H_1$ in the basis
$\cB$ is tridiagonal, Hermitian, and has zero diagonal entries.
The only nonzero entries of $\sf H_1$ are given by
\begin{equation}
\fl
  ({\sf H_1})_{s+1,s}=\left(
  \prod_{l=1}^{N}\prod_{j_l=0}^{n_l-1}(\cN_l+n_ls+j_l+1)\cdot
  \prod_{k=1}^{M}\prod_{i_k=0}^{m_k-1}(\cM_k-m_ks-i_k)\right)^{\!\frac
  12}=({\sf H_1})_{s,s+1}
\end{equation}
with $s=0,1,\dots,r-1$.

A number of immediate conclusions can be drawn
from the structure of the matrix $\sf H_1$. We note in the first place that
$H_1$ acts irreducibly in $\cS^{\cN}_{\cM}$. This implies that $H_1$
is functionally independent from $H_0$, $A_l$ and $B_k$ for all $k,l$. Since
$x^ny^m$ is annihilated by all the $A_l$ and $B_k$ but not by $H_0$, the
latter operator is functionally independent from the former ones.
Since all the $A_l$ and $B_k$ are clearly functionally independent,
the preceding arguments show the functional independence of all the
operators in the set~(\ref{commset}) and finish the proof of the complete
integrability claimed at the end of section~2.

Secondly, the eigenvalues of $\sf H_1$ are real, simple,
and symmetrically distributed around zero. Indeed, let
$\de_s(E)$, $1\le s\le r+1$, be the $s$-th principal minor of the matrix
$E\,\id-{\sf H_1}$. A straightforward computation shows that
$\de_s(E)$ satisfies the three-term recursion relation
\begin{equation}\label{rr}
  \de_{s+1}(E)=E\,\de_s(E)-h_s^2\de_{s-1}(E)\qquad s\ge 1
\end{equation}
where $h_s=({\sf H_1})_{s,s-1}$ and $\de_1(E)=E$, $\de_0(E)=1$.
Since $h_s^2>0$ for $1\le s\le r$, lemma~1 (section 1.8) of~\cite{Ar64}
implies that the roots of the characteristic polynomial $\de_{r+1}(E)$
are real and simple. They are also symmetrically distributed around zero,
for $\de_{r+1}(E)$ contains only either even or odd powers of $E$ on account
of the form of the recursion relation~(\ref{rr}).\\

\nid{\it Remark.} The model~(\ref{sys}) admits certain solvable (and
in some cases integrable) generalizations.  In the first place, we can
replace $g H_1$ by any polynomial in $H_1$ with real coefficients.
Clearly, the resulting Hamiltonian is completely integrable, since it
commutes with the operators~(\ref{commset}).  Moreover, this
generalized Hamiltonian preserves the finite-dimensional subspaces
$\cS^\cN_\cM$ and is therefore exactly solvable.

More generally, one
can consider a perturbation Hamiltonian $H'_1$ consisting of a finite sum of terms
of the form~(\ref{H1}), provided that the exponents in each term satisfy
condition~(\ref{constraint}). In other words,
\begin{equation}
  H'_1 = \sum_{(n,m)\in S}g_{nm}\left(\prod_{k=1}^M (b_k^{\dagger})^{m_k}\cdot
        \prod_{l=1}^N a_l^{n_l}
    +
        \prod_{l=1}^N (a_l^{\dagger})^{n_l}\cdot
        \prod_{k=1}^M b_k^{m_k}\right)
  \label{h1gen}
\end{equation}
where $S$ is any finite subset of the set of all pairs
$(n,m)\in\ZZ_+^N\times\ZZ_+^M$ satisfying the
condition~(\ref{constraint}).  Note that $H_0$ still commutes with $H'_1$,
since it commutes separately with each term in the sum~(\ref{h1gen}).
Thus the spectrum of $H$ can be computed by diagonalizing the new
perturbation Hamiltonian $H'_1$ in the eigenspaces of $H_0$, which are
finite-dimensional by equation~(\ref{E0}).  Hence this more general
model is also exactly solvable.  Note, however, that now to each term in the sum~(\ref{h1gen}) there correspond two families of operators
$A_l\equiv A_l(n)$ and $B_k\equiv B_k(m)$ which
do not necessarily commute with the remaining terms in $H'_1$.  Hence,
this model need not be completely integrable for general values of the
coupling constants $g_{nm}$.
\section{Quasi-exact solvability of the reduced Hamiltonians}

In the previous section we have shown that the computation of the
spectrum of the Hamiltonian~(\ref{sys}) is equivalent to the
diagonalization of a family of tridiagonal Hermitian matrices with
zero diagonal elements.  For small values of
$r+1=\dim\cS^{\cN}_{\cM}$, the eigenvalues of the matrix $\sf H_1$
representing $H_1$ can be easily computed in closed form.  In many
situations of physical interest, however, the number of photons of
each frequency involved can actually be very large.  This implies (see
equation~(\ref{r})) that the parameter $r$ can also take large values.
Although the spectrum of $\sf H_1$ can be determined numerically for
fixed values of the parameters $\cN_l$, $\cM_k$ labelling the space
$\cS^{\cN}_{\cM}$, one is often interested in the behavior of its
eigenvalues as a function of $\cN_l$, $\cM_k$ (and thus $r$).  In the
case of second-harmonic generation, i.e. $N=M=m_1=1,n_1=2$,
\'{A}lvarez and \'{A}lvarez-Estrada~\cite{AA95} derived
asymptotic formulae for the eigenvalues of $\sf H_1$ by performing a
semiclassical analysis of the second-order ordinary differential
equation (ODE) obtained by restricting the perturbation Hamiltonian $H_1$
to $\cS^{\cN_1}_{\cM_1}=\cS^{\ep}_r$, where $\ep=0,1$.  It was also shown
in that paper that the restriction of $H_1$ to $\cS^{\ep}_r$ is in
fact equivalent to a QES operator in a single variable, in agreement
with previous results due to Zaslavskii~\cite{Za90}.

In this section we show that the restriction of our general
perturbation $H_1$ to an eigenspace $\cS^{\cN}_{\cM}$ is also
equivalent to a QES differential operator in a single variable.
Besides the purely mathematical interest of this result, the knowledge
of explicit expressions for the ordinary differential operator of the
restriction $H_1|_{\cS^{\cN}_{\cM}}$ is precisely the starting point
to apply asymptotic techniques, which are currently being
developed for third- and higher-order ODE's.

Consider the action of $H_1$ on an element
$x^\cN y^\cM P(\ze)\in\cS^{\cN}_{\cM}$
\begin{equation}
\fl
  H_1\left(x^\cN y^\cM P(\ze)\right)=
  \prod_{k=1}^M y_k^{m_k+\cM_k}\cdot
  \prod_{l=1}^N
  \pa_{x_l}^{n_l}(x_{l}^{\cN_l} P(\ze))
  +
  \prod_{l=1}^N x_l^{n_l+\cN_l}\cdot
  \prod_{k=1}^M\pa_{y_k}^{m_k}(y_k^{\cM_k} P(\ze))
\end{equation}
where $P$ is a polynomial of degree at most $r$.
Since
\begin{eqnarray}
  \pa_{x_l}^{n_l}(x_{l}^{\cN_l}P(\ze))
  =
  x_{l}^{\cN_l-n_l} \prod_{j_l=0}^{n_l-1}
  (\cN_l-j_l+n_l\,\ze\pa_\ze)P(\ze) \\
  \pa_{y_k}^{m_k}(y_k^{\cM_k} P(\ze))
  =
  y_{k}^{\cM_k-m_k}\prod_{i_k=0}^{m_k-1}
  (\cM_k-i_k-m_{k}\,\ze\pa_\ze)P(\ze)
\end{eqnarray}
it follows that
\begin{equation}\label{actions}
  H_1\left(x^\cN y^\cM P(\ze)\right)
  =
  x^\cN y^\cM H_{1,\rm red}\,P(\ze)
\end{equation}
where the reduced Hamiltonian is given by
\begin{equation}
  H_{1,\rm red}
  =
  \frac 1\ze\prod_{l=1}^N\prod_{j_l=0}^{n_l-1}
  (\cN_l-j_l+n_l\,\ze\pa_\ze)
  +
  \ze\prod_{k=1}^M
  \prod_{i_k=0}^{m_k-1}(\cM_k-i_k-m_{k}\,\ze\pa_\ze).
  \label{H1eff}
\end{equation}
Thus the action of $H_1|_{\cS^{\cN}_{\cM}}$ is equivalent to that of
$H_{1,\rm red}$ on the space $\cP_r$ of polynomials in $\ze$ of degree
at most $r$. Note in particular that the invariance of $\cS^\cN_\cM$ under
$H_1$ implies that $\cP_r$ is invariant under $H_{1,\rm red}$ (this
fact can be also verified directly from equation~(\ref{H1eff})
using~(\ref{r}) and (\ref{restric}); see the discussion below).

Let $d=\max(\sum_{l=1}^N n_l,\sum_{k=1}^M m_k)$ be the
order of the differential operator $H_{1,\rm red}$.  If $d\le r$, a
well-known theorem due to Turbiner~\cite{Tu92} (see~\cite{FK98} for a
simplified proof) states that the invariance of the polynomial space
$\cP_r$ under $H_{1,\rm red}$ implies that this operator can be
written as a polynomial of degree $d$ in the generators of the
realization of $\fsl_2$ spanned by
\begin{equation}\label{Js}
  J^+=\ze^2\pa_\ze-r\ze
  \qquad
  J^0=\ze\pa_\ze-{r\over 2}
  \qquad
  J^-=\pa_\ze.
\end{equation}
If $d>r$, Turbiner's theorem only guarantees that the $r$-th order part
of $H_{1,\rm red}$ is a polynomial of degree $d$ in the operators~(\ref{Js}).
In fact, even if $d>r$ the reduced Hamiltonian $H_{1,\rm red}$ can be
written as a polynomial of degree $d$ in the generators~(\ref{Js}).
Indeed, choose $k'\in\{1,\dots,M\}$ and $l'\in\{1,\dots,N\}$
such that (see equations~(\ref{r}) and~(\ref{restric}))
\begin{equation}
  r = \left[\frac{\cM_{k'}}{m_{k'}}\right]
  \qquad
  \mbox{and}
  \qquad
  \cN_{l'} < n_{l'}.
\end{equation}
Since all the factors in the products of the expression~(\ref{H1eff})
commute, it follows that
\begin{eqnarray}
\fl
  H_{1,\rm red} =
  n^n \, J^-\cdot
  \prod_{j_{l'}=0,\,j_{l'}\ne\cN_{l'}}^{n_{l'}-1}
  \left(J^0+\frac{\cN_{l'}-j_{l'}}{n_{l'}}+\frac r2\right)\cdot
  \prod_{l=1,\,l\ne l'}^N
  \prod_{j_l=0}^{n_l-1}
  \left(J^0+\frac{\cN_{l}-j_{l}}{n_{l}}+\frac r2\right)
  \nonumber\\
  {}+
  (-m)^m\,J^+ \cdot
  \prod_{i_{k'}=0,\,i_{k'} \ne \cM_{k'}\bmod m_{k'}}^{m_{k'}-1}
  \left(J^0+\frac{i_{k'}-\cM_{k'}}{m_{k'}}+\frac r2\right)
  \nonumber\\
  \qquad
  \cdot
  \prod_{k=1,\, k\ne k'}^M\prod_{i_k=0}^{m_k-1}
  \left(J^0+\frac{i_{k}-\cM_{k}}{m_{k}}+\frac r2\right).
  \label{eq:h1red}
\end{eqnarray}
This general expression of the differential operator $H_{1,\rm red}$
as a polynomial in the generators of the $\fsl_2$ algebra is of course
not unique, and can be written in different forms using the
commutation relations of the generators~(\ref{Js}).  We conclude this
section by showing explicitly the physically most important particular
cases, in which the general expression~(\ref{eq:h1red})
simplifies considerably.

\nid{\bf Example 1} Consider the problem of $n$-th harmonic
generation, in which $N=M=m_1=1$, $n_1=n$ (and therefore the
multiindices reduce to ordinary indices). The eigenspaces of $H_0$ are
$\cS_r^\ep=x^\ep y^r\cP(\ze)$, where $r\in\ZZ_+$, $\ep=0,\dots,n-1$,
and $\ze=x^n/y$. The corresponding expression of
$H_{1,\rm red}$ reads
\begin{equation}
  H_{1,\rm red}
  =
  n^n J^-\,\prod_{j=0,\, j\ne\ep}^{n-1}
  \left(J^0+\frac{\ep-j}n+\frac r2\right)-J^+\,.
\end{equation}
This expression can be easily shown to be equivalent to the
expressions obtained in~\cite{AA95} for second harmonic generation ($n=2$)
and in~\cite{AA01} for third-harmonic generation ($n=3$) and $\ep=0$
(incidentally, in~\cite{AA01} the cases $\ep=1$ and~$2$ were not considered).

\nid{\bf Example 2} In the case of a multiple photon cascade,
$M=m_1=n_1=\cdots=n_N=1$, the eigenspaces of $H_0$ and $A_l$,
$l=1,\dots,N-1$, are $\cS^\cN_r=x^\cN y^r\cP(\ze)$, where
$\cN_{l'}=0$ for some $l'\in\{1,\dots,N\}$, $r\in\ZZ_+$,
and $\ze=(x_{1}\cdots x_{N})/y$.
The expression of $H_{1,\rm red}$ in terms of the
generators~(\ref{Js}) is
\begin{equation}
  H_{1,\rm red}
  =
  J^-\,\prod_{l=1,\, l\ne l'}^{N}
  \left(J^0+\cN_l+\frac r2\right)-J^+\,.
\end{equation}
\section{Summary}
In this paper we have established a common framework to deal with a
large class of processes in nonlinear optics including the problems of
$n$-th harmonic generation and multiple photon cascades.  The
distinguishing feature of these models from the physical point of view
is the condition of energy conservation, which is the key mathematical
condition to prove their complete integrability and the exact
solvability.  By using the Bargmann representation, we have been able
to provide an explicit description of the common unperturbed
eigenspaces, as well as equally explicit expressions of the action of
the perturbation Hamiltonian in these unperturbed eigenspaces.  We
have also derived some general properties of the spectrum of the
restriction of the perturbation $H_{1}$ to each unperturbed
eigenspace, such as the nondegeneracy and symmetric distribution of
the perturbation energies around zero.  The link between these finite
matrix representations and the corresponding continuous QES system is
most conveniently established by the introduction of a new
``projective'' coordinate, which is a quotient of powers of the
Bargmann variables that describe the different oscillators (and
physically carries information on the phase difference among the
oscillators).  Furthermore, we have been able to give explicit
expressions of the reduced QES Hamiltonians as polynomials in the
generators of the standard QES realization of $\fsl_{2}$ by
first-order differential operators.

\end{document}